\documentclass[prl,twocolumn,showpacs,groupedaddress]{revtex4}
\usepackage{graphicx}
\begin{document}

\newcommand{\ket}[1]{|#1\rangle}
\newcommand{\bra}[1]{\langle#1|}

\def\aL{ \hat{a}^{\dag}_{L} }
\def\aR{ \hat{a}^{\dag}_{R} }
\def\bL{ \hat{b}^{\dag}_{L} }
\def\bR{ \hat{b}^{\dag}_{R} }

\title{Adiabatic evolution under quantum control}
\author{ W. Wang, S. C. Hou, and  X. X. Yi}
\affiliation{School of Physics and Optoelectronic Technology, Dalian
University of Technology, Dalian 116024, China}

\date{\today}

\begin{abstract}
One of the difficulties in adiabatic quantum computation is the
limit on the computation time. Here we propose two schemes to
speed-up the adiabatic evolution. To apply this controlled adiabatic
evolution to adiabatic quantum computation, we design one of the
schemes without any prior knowledge of the instantaneous eigenstates
of the final Hamiltonian. Whereas in another scheme, the control is
constructed with the instantaneous eigenstate that is the target
state of the control. As an illustration, we study a two-level
system driven by  a time-dependent magnetic field  under the
control. The physics behind the control scheme  is explained.
\end{abstract}

 \maketitle


Adiabatic quantum computation (AQC) was first proposed in
2000\cite{Farhi01} as a means to solve NP-complete problems. It is
polynomially equivalent to conventional (gate model) quantum
computation\cite{adkllr04} and  possesses some degree of fault
tolerance. AQC can be described by the following Hamiltonian,
 \begin{equation}
 H_0(t) = [1-\lambda(t)]H_i + \lambda(t) H_f, \label{HS}
 \end{equation}
where the quantum system govern by $H_0(t)$ evolves slowly with time
$t$ and remains in its ground state as $\lambda(t)$ changes
monotonically from 0 to 1 within a time $T$. The initial Hamiltonian
$H_i$ is assumed to have an easily accessible ground state into
which the system is initialized, while the ground state of the final
Hamiltonian $H_f$ encodes a problem's solution. In order to reach
the final ground state with high fidelity, the adiabatic theorem
requires $T \propto (\Delta E_{\rm min})^{-\delta}$, where $\Delta
E_{\rm min}$ is the minimum gap between the two lowest energy
 eigenstates of $H_0(t)$. The power $\delta$ can be 1,
2, or possibly some other number depending on the functional form of
$\lambda(t)$ and the distribution of the higher energy levels
\cite{Farhi01,schaller06,lidar08}. This leads to  a limit  on the
computation time $T$,  which holds true for adiabatic quantum
computation {\it without quantum control}, thus it is natural to ask
whether one can use quantum control  to speed up AQC?

Quantum control\cite{lloyd00,viola98,ramakrishna95,schirmer01,
wiseman94,mancini98,belavkin99,doherty99,carvalho07,yi08} is about
the application of classical and modern control strategy  to quantum
systems. It has generated increasing interest in the last few years
due to its potential applications in
metrology\cite{wiseman95,berry01},
communications\cite{geremia04a,jacobs07} and other technologies
\cite{ahn02,hopkins03,geremia04b,sarovar04,steck04}, as well as its
theoretical interest on its own. Several approaches to the control
of a quantum system have been proposed in the past decade, which can
be divided into coherent (unitary) and incoherent  (non-unitary)
control, according to how the controls enter the dynamics. In the
coherent control scheme, the controls enter the dynamics through the
system Hamiltonian. It affects the time evolution of the system
state, but not its spectrum, i.e., the eigenvalues of the target
density matrix $\rho_f$ remain unchanged in the dynamics. In the
incoherent control scheme, an auxiliary system, called probe, is
introduced to manipulate the target system through their mutual
interaction. This incoherent control scheme is of relevance whenever
the system dynamics can not be directly accessed and  provides a
non-unitary evolution for the quantum system. This breaks the
limitation for the coherent control mentioned above. Among these
quantum control strategies, Lyapunov control plays an important role
in quantum control theory. Lyapunov functions, originally used in
control to analyze the stability of the control system, have formed
the basis for new control design. Several papers have be published
recently to discuss the application of  Lyapunov control to quantum
systems\cite{vettori02,ferrante02,grivopoulos03,
mirrahimi04,mirrahimi05,altafini07,wang08,yi09}. Although the basic
mathematical formulism for Lyapunov control is well established,
many questions remain when one considers its applications in quantum
information processing, for instance, whether one can use quantum
Lyapunov control to improve  the adiabatic evolution, and
consequently  minimize the computation time?

In this paper, we shall address this issue by using a two-level
model with Lyapunov control. Since the two lowest levels are
important for AQC, this model to some extent  can good quantify  the
AQC under control. Two control schemes are proposed which correspond
to two different choices of Lyapunov function. By numerical
simulation, we find that these schemes work well.  The paper is
organized as follows. In Sec.{\rm II}, we present a general
formalism for the Lyapunov control, two Lyapunov functions which
will give two control schemes are constructed, then we use these
schemes to manipulate  a two-level system in Sec.{\rm III}.
Conclusion and discussions are presented in Sec.{\rm IV}.

{\it General formalism.---} Let us start by focusing on the
Hamiltonian $H_0(t)$ in Eq.(\ref{HS}). We denote the instantaneous
eigenstates and the corresponding eigenvalues of $H_0(t)$ by
$|n(t)\rangle$ and $E_n(t)$, respectively. AQC is designed to take
advantage of the adiabatic theorem, it works by evolving a system
from the accessible ground state of an initial Hamiltonian $H_i$ to
the ground state of a final Hamiltonian $H_f.$  For the system to
remain in its ground state, the adiabatic theorem requires that the
system Hamiltonian changes very slowly and there is an energy gap
between the ground state and the others.  The  goal of this paper is
to manipulate the system such that it remains in  one of its
eigenstates. To meet the requirement of AQC, any information about
the instantaneous eigenstates of $H_f$ is forbidden to use in the
control, but a prior knowledge regarding the system Hamiltonian is
allowed. This motivates us to propose the first scheme in the
following.  Beside the interests in AQC, controlling a quantum
system to a target state is interesting on its own. As the target
state is known, the information about the target state can be used
in the control design. This is the goal of the second scheme
discussed in  this paper.

{\it Scheme A.}--- We aim to design a control scheme that can
manipulate a system to remain in its instantaneous ground state
without any prior knowledge of its instantaneous eigenstates of
$H_f$. To this end, we introduce  control operators $H_{cj} \
(j=1,2,3,...)$ and require that $[H_0,H_{cj}]=0$ for any $j.$  The
control operators $H_{cj}$ enter the system through  control field
$f_j(t).$ The total Hamiltonian of the system is then written,
\begin{equation}
H=H_0+\sum_j f_j(t)H_{cj},
\end{equation}
where the control field $f_j(t)$ can be established by Lyapunov
control theory. Define a function $V(t)$ as
\begin{equation}
V(t)=|\langle\psi(t)|X|\psi(t)\rangle|^2,
\end{equation}
with $X$ being a time-independent Hermitian operator, we find
$V(t)\geq 0$ and
\begin{equation}
\dot{V}(t)=\frac{2i}{\hbar}\langle \psi(t)|X|\psi(t)\rangle \langle
\psi(t)|[H,X]|\psi(t)\rangle.
\end{equation}
Here and hereafter $|\psi(t)\rangle$ denotes the state of the system
that satisfy $i\hbar\frac{\partial}{\partial t}|\psi(t)\rangle
=H|\psi(t)\rangle.$  We now show how to establish the control field
$f_j(t)$. Lyapunov control theory tells that for $V(t)$ to be a
Lyapunov function, $V(t)$ has to satisfy, $V\geq 0$ and $\dot{V}\leq
0.$ So, if we choose the control field $f_j(t)$ as
\begin{eqnarray}
f_j(t)&=&i\langle\psi(t)|X|\psi(t)\rangle\langle\psi(t)|[H_{cj},
X]|\psi(t)\rangle,\ \ \mbox{for}\ \  j\neq j_0, \nonumber\\
f_{j_0}(t)&=&-\frac{\langle\psi(t)|[H_0,
X]|\psi(t)\rangle}{\langle\psi(t)|[H_{cj_0}, X]|\psi(t)\rangle},\ \
\mbox{for}\ \  j=j_0, \label{f1}
\end{eqnarray}
then $\dot{V}\leq 0.$ Here $j_0$ was specified to satisfy $\langle
\psi|[H_{cj_0},X]|\psi\rangle\neq 0.$ Clearly $f_j(t)$ is a
time-dependent real number, thus the total Hamiltonian $H$ is
Hermitian. The key point of this control scheme is the choice of
operator $X,$ it dominates the success rate of the control and show
the merit in this scheme. From the control design of the scheme, we
find that the instantaneous eigenstates do not enter the control,
this mean that by the present control strategy, we can manipulate
the quantum system to evolve along one of the eigenstates without
any knowledge of its instantaneous eigenstates. This is exactly what
we want in AQC. To see that this control scheme indeed works, we
note that $[H_0,H_{cj}]=0,$ indicating the control $H_{cj}$ itself
can not induce population transfer among the instantaneous
eigenstates. Suppose that the system is initially prepared in its
ground state $|0(t=0)\rangle$ and $H_{cj}=H_0$, Eq.(\ref{f1}) yields
$f_{j_0}(t=0)=-1$ and $f_j(t=0)=0, $ $(j=1,2,3,...,j\neq j_0),$
regardless of what $X$ takes. When $|\psi(t)\rangle$ deviates from
$|0(t)\rangle$, $f_j(t)$ can be very large depending on operator
$X$. The Lyapunov control will then render the system nonlinear, and
this nonlinear effect would bring the state $|\psi(t)\rangle$ back
to $|0(t)\rangle.$ We note that $\frac{\partial H_{cj}}{\partial t}$
(as well as $\frac{\partial H_0}{\partial t})$ can derive the system
from one instantaneous eigenstate to the others, this together with
the control keep the system in the instantaneous eigenstate into
which the system was initially prepared. We will demonstrate this
point through an example in detail later.

{\it Scheme B.}---For general control problem, the target state is
known, we then can use the target state to design a Lyapunov
function and establish the control field $f_j(t).$  Suppose that the
target state is the zeroth instantaneous eigenstate $|0(t)\rangle$
of $H_0(t)$, define
\begin{equation}
V(|0(t)\rangle, |\psi(t)\rangle)=1-|\langle 0(t)|\psi(t)\rangle|^2.
\label{lyaf2}
\end{equation}
Clearly $V\geq 0$ with equality only if
$|0(t)\rangle=\ket{\psi(t)}.$ To see Eq.(\ref{lyaf2}) indeed defines
a Lyapunov function, we calculate the time derivative of $V$
as($H=H_0+\sum_jf_j(t)H_{cj}^{\prime}$ in this scheme),
\begin{eqnarray}
\dot{V}&=&-\frac{i}{\hbar}
\sum_jf_j(t)(\langle\psi(t)|H_{cj}^{\prime}|0(t)\rangle\langle
0(t)|\psi(t)\rangle-c.c.)\nonumber\\
&-&2\mbox{Re}(\langle\dot{0}(t)|\psi(t)\rangle\langle\psi(t)|0(t)\rangle).
\end{eqnarray}
Obviously,
\begin{eqnarray}
f_j(t)=&&-2\mbox{Im}(\langle\psi(t)|H_{cj}^{\prime}|0(t)\rangle\langle
0(t)|\psi(t)\rangle), \ \mbox{for}\ j\neq j_0,\nonumber\\
f_{j_0}(t)=&&-\frac{\hbar\mbox{Re}(\langle
\dot{0}(t)|\psi(t)\rangle\langle\psi(t)|0(t)\rangle)}{\mbox{Im}(\langle
0(t)|H_{cj_0}^{\prime}|\psi(t)\rangle\langle\psi|0(t)\rangle)},  \
\mbox{for} \ j=j_0,\label{f2}
\end{eqnarray}
guarantee that $\dot{V}\leq 0.$ Again, $j_0$ was selected to satisfy
$\mbox{Im}(\langle
0(t)|H_{cj_0}^{\prime}|\psi(t)\rangle\langle\psi|0(t)\rangle)\neq
0.$  Hence, the evolution of the system with Lyapunov control can be
described by the following nonlinear autonomous equations,
\begin{eqnarray}
&&i\hbar\frac{\partial}{\partial
t}|\psi(t)\rangle=H|\psi(t)\rangle,\nonumber\\
&&H=H_0+\sum_jf_j(t)H_{cj}^{\prime},\nonumber\\
&&H_0|0(t)\rangle=E_0|0(t)\rangle,\label{deq1}
\end{eqnarray}
where $f_j(t)$ $(j=1,2,3,...)$ are given by Eq.(\ref{f2}).
 Note that $[H_{cj}^{\prime},H_0]=0$ is not required in this
scheme. The difference between the present scheme and those in the
literature\cite{wang08} is the target state. In earlier studies, the
target state $|\phi_d\rangle$ is either time-independent or
time-dependent via $i\hbar\frac{\partial}{\partial
t}|\phi_d\rangle=H_0|\phi_d\rangle$, whereas in our scheme, the
target state is one of the instantaneous eigenstate of $H_0$. The
time derivative of the instantaneous eigenstate plays  an important
role in the control, leading to  totally different control fields
$f_j(t)$ in Eq.(\ref{f2}). Note that the choice of $f_j(t)$ in
Eqs.(\ref{f1})  and (\ref{f2}) are not unique. In fact, when there
is only one control operator $H_c$, the control field $f(t)$ in
Eq.(\ref{f1}) can be chosen as,
\begin{eqnarray}
f(t)&=&i\langle\psi(t)|X|\psi(t)\rangle\langle\psi(t)|[H_c,
X]|\psi(t)\rangle \nonumber\\
&-&\frac{\langle\psi(t)|[H_0,
X]|\psi(t)\rangle}{\langle\psi(t)|[H_c,
X]|\psi(t)\rangle},\label{f11}
\end{eqnarray}
provided $f(t)$ is a finite number.  This is exactly the case in the
example that we will illustrate below.

{\it Example.---} As an illustration of the Lyapunov  control
scheme, we discuss below a two-level  system driven by a
time-dependent magnetic field. The Hamiltonian that describes such a
system can be written as
\begin{eqnarray} \label{spin}
H_0(t)&=&\mu \vec{B}\cdot \vec{\sigma}\\
&=&\mu
B_0(\sin\theta\cos\phi\sigma_x+\sin\theta\sin\phi\sigma_y
+\cos\theta\sigma_z),\nonumber
\end{eqnarray}
where $\sigma_x, \sigma_y$ and $\sigma_z$ are the Pauli matrices,
$B_0$ is the amplitude of the classical field. $\theta$ is specified
to be a constant here, while $\phi$ depends on time $t$ through
$\phi=\omega t$\ \ ($\omega$, constant). In comparison with
Eq.(\ref{HS}), $H_i$ takes $\mu
B_0(\sin\theta\sigma_x+\cos\theta\sigma_z),$ while $H_f=\mu
B_0(\sin\theta\sigma_y+\cos\theta \sigma_z)$. Although in this model
we can not found an analytical $\lambda(t)$ by which we write
Eq.(\ref{spin}) in the form of Eq.(\ref{HS}), the model
Eq.(\ref{spin}) can be obviously mapped into Eq.(\ref{HS}). It is
well know that the instantaneous eigenstates and the corresponding
eigenvalues of $H_0$ are
$|E_+(t)\rangle=\cos\frac{\theta}{2}e^{i\phi}|\uparrow\rangle
+\sin\frac{\theta}{2}|\downarrow\rangle,$
$|E_-(t)\rangle=-\sin\frac{\theta}{2}e^{i\phi}|\uparrow\rangle
+\cos\frac{\theta}{2}|\downarrow\rangle$ and $E_{\pm}=\pm\mu B_0$,
respectively.
\begin{figure}
\includegraphics*[width=0.8\columnwidth,height=0.55\columnwidth]{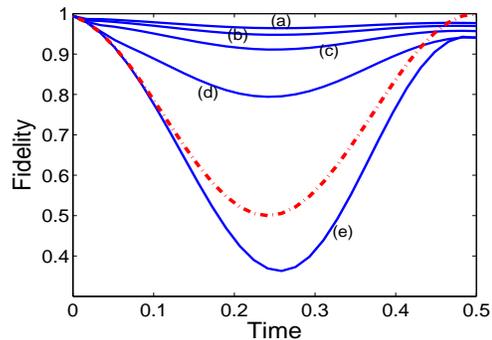}
\caption{(Color online) Fidelity defined by $F(t)=|\langle
E_-(t)|\psi(t)\rangle|^2$ as a function of time. The system is
initially prepared in $|E_-(t=0)\rangle.$ (a)-(e) correspond to
different ratio $R$. From (a) to (e), $R=12,9,6,3,0$, respectively.
By contrast, the result without control is shown by the dashed line.
We have set $\mu B_0=1$ for this plot. The time is shown in units of
$1/(\mu B_0)$, the other parameters chosen are $\omega=4$ (in units
of $\mu B_0$) and $\theta=\frac{\pi}{4}.$ Note $\omega=4\mu B_0,$
the adiabaticity breaks for the system without control.}
\label{fig1}
\end{figure}
In the  absence of Lyapunov control, it is required that $\omega\ll
\mu B_0$ for the system to evolve adiabatically. We now show that
the system can evolve along one of its instantaneous eigenstates,
e.g. $|E_-(t)\rangle$,  under the Lyapunov control even if $\omega
\ge \mu B_0.$  In the following, a fidelity defined by
\begin{equation}
F(t)=|\langle E_-(t)|\psi(t)\rangle|^2,
\end{equation}
will be used to measure the effectiveness of the control. The
dependence of $F(t)$ on time $t$ and the control operator
$H_c^{\prime}$ as well as $X$ are shown. The results show that these
schemes work good with properly chosen $X$ and $H_c^{\prime}$.

We first consider the scheme A, where no information about
instantaneous eigenstates enter the control. Choose
$X=\sigma_x+R\sigma_z$ with a rate $R$, the dynamics of the
two-level  system under control is governed by
\begin{eqnarray}
i\hbar\frac{\partial}{\partial
t}|\psi(t)\rangle&=&H|\psi(t)\rangle,\nonumber\\
H&=&\mu\vec{B}\cdot \vec{\sigma}+f(t)H_c,\nonumber\\
f(t)&=&i\langle\psi(t)|X|\psi(t)\rangle\langle\psi(t)|[H_c,
X]|\psi(t)\rangle \nonumber\\
&-&\mu B_0. \label{deq2}
\end{eqnarray}
Here $H_c$ was specified to be
$H_c=(\sin\theta\cos\phi\sigma_x+\sin\theta\sin\phi\sigma_y+\cos\theta\sigma_z)$
in order to satisfy $[H_c,H_0]=0.$ We have perform extensive
numerical simulations for the dynamics Eq.(\ref{deq2}), selected
results are plotted in Fig. \ref{fig1}. Two observations can be
found from Fig.\ref{fig1}. (1) The scheme works only for special
$X$. For some choices of  $X$, the control favors the adiabatic
evolution, but for the other $X$, the control spoils the
adiabaticity of the evolution. (2) For the specified $X$,
$X=\sigma_x+R\sigma_z$, the larger the ratio $R$ is, the better the
fidelity for the system in $|E_-(t)\rangle$.

\begin{figure}
\includegraphics*[width=0.8\columnwidth,height=0.55\columnwidth]{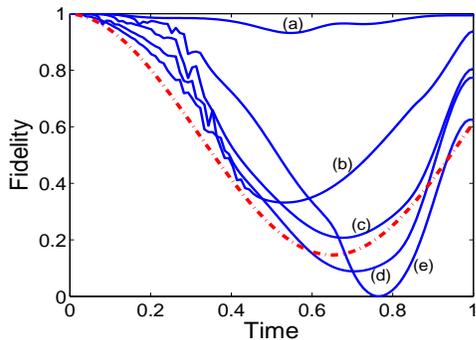}
\caption{(Color online)Fidelity versus time $t$ given by the scheme
B. The fidelity was defined as the same as in Fig.\ref{fig1}. From
(a) to (e), the ratio $r$ takes $0,2,4,6,8,$ respectively. Dashed
line shows the result without control. All quantities are chosen and
plotted in the same units as in Fig.\ref{fig1}.} \label{fig2}
\end{figure}

In the scheme B, a prior knowledge of the instantaneous eigenstates
is known and allowed to use to design the control field $f(t)$.
Choose $|E_-(t)\rangle$ as the target state, the dynamics of the
system is governed by Eq.(\ref{deq1}) with replacing $|0(t)\rangle$
by $|E_-(t)\rangle$. We show in Fig.\ref{fig2} the fidelity as a
function of time with $H_c^{\prime}=\sigma_z+ r\sigma_x.$
Similar to the results given in the scheme A, the choice of the
control operator $H_c^{\prime}$ dominates  the effectiveness of the
control. Not all control operator $H_c^{\prime}$ can help the
adiabatic evolution. The difference between these two scheme is that
the scheme B can reach fidelity 1 at the final stage, this may
depend on the example demonstrated.
\begin{figure}
\includegraphics*[width=0.8\columnwidth,height=0.55\columnwidth]{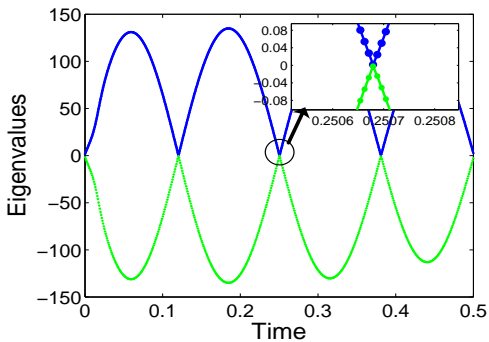}
\caption{(Color online)Eigenvalues of Hamiltonian $H$ as a function
of time. The parameters chosen are the same as in
Fig.\ref{fig1}-(a), indicating that the system remains well in one
of the instantaneous eigenstates under control. The inset shows an
enlarged level anti-crossing point.} \label{fig3}
\end{figure}

For linear quantum system, the population transfer from one
instantaneous eigenstate to the others depends on the energy gap and
the operator that induces the transfer. The energy gap separates the
instantaneous eigenstates and ensures the adiabaticity of the
evolution. The larger the energy gap is, the higher the probability
with which the system remains in the instantaneous eigenstate.  In
the scheme A, the operator that induces the population transfer is
$\delta H=(\frac{\partial H_0}{\partial t}+f(t)\frac{\partial
H_{c}}{\partial t}),$ small $\delta H$ would benefits the
adiabaticity of the system. For nonlinear system, however, the
nonlinearity plays an important role in addition to the factors for
linear system mentioned above. All these factors together lead to
the observed features and can be used to understand the physics
behind the features.

To make the explanation clear, we plot the eigenvalues for the total
Hamiltonian in scheme A as a function of time in Fig.\ref{fig3}. The
same parameters as in Fig.\ref{fig1}-(a) are chosen, which ensure
that  the state of the system remains in the instantaneous
eigenstate $|E_-(t)\rangle$. Recall that the energy gap between two
instantaneous eigenstates is 2 (in units of $\mu B_0$) for $H_0$, we
find from Fig.\ref{fig3} that the energy gap is intensively enlarged
at most times by the control. Nevertheless, the enlarged energy gap
itself can not explain why the system remains well in the
instantaneous eigenstate, because the energy gap is very small at
the level anti-crossing points, this indicates that the nonlinearity
must play an important role in the dynamics. Indeed our numerical
calculation shows that at these point, the nonlinear coefficient
defined by $|\langle E_-(t)|f(t) H_c|E_-(t)\rangle|$ is much larger
than the tunneling coefficient $|\langle E_-(t)|\delta
H|E_+(t)\rangle|,$ leading to the observed feature reminiscent of
self-trapping in nonlinear system.


To sum up, we have proposed two schemes to speed-up the adiabatic
evolution. In the scheme A, the control has been designed without
any information about the instantaneous eigenstates of the final
Hamiltonian, hence this scheme can be used in the adiabatic quantum
computation. The scheme B has been proposed using the instantaneous
eigenstate of the system Hamiltonian. This scheme is applicable  to
control a quantum system when the target state is known. To show how
the schemes work, we have presented an example consisting of a
two-level system in a rotating magnetic field. The fidelity to
quantify the effectiveness of the scheme was calculated and
discussed. The results show that the fidelity  depends sharply on
the choice of $X$ and the control operator $H_c^{\prime}$. The
physics behind the schemes is revealed, which can be understood as
an effect combining  nonlinear effects and the broadened energy gap.

This work was supported by  NSF of China under grant Nos. 10775023
and 10935010.


\begin{references}

\bibitem{Farhi01} E. Farhi, J. Goldstone, S. Gutmann,
J. Lapan, A. Lundgren, and D. Preda,
Science {\bf 292}, 472 (2001).

\bibitem{adkllr04}
D.~Aharonov, W.~van Dam, J.~Kempe, Z.~Landau, S.~Lloyd, and
O.~Regev, {\em Proc. 45th FOCS}, 42 (2004).


\bibitem{schaller06} G. Schaller, S. Mostame, and  R. Sch\"utzhold
Phys. Rev. A {\bf 73}, 062307 (2006).

\bibitem{lidar08} D.A. Lidar, A.T. Rezakhani, and A. Hamma, eprint arXiv:0808.2697.


\bibitem{lloyd00} S. Lloyd, Phys. Rev. A {\bf 62}, 022108(2000).

\bibitem{viola98} L. Viola and S. Lloyd, Phys. Rev. A {\bf 58},
2733(1998); L. Viola, E. Knill, and S. Lloyd, Phys. Rev. Lett {\bf
82}, 2417(1999).


\bibitem{ramakrishna95} V. Ramakrishna, M. V. Salapaka, M. Dahleh, H. Rabitz,
A. Peirce,   Phys. Rev. A {\bf 51}, 960(1995).

\bibitem{schirmer01} S. G. Schirmer, H. Fu, and A. I. Solomon, Phys.
Rev. A {\bf 63}, 063410(2001); H. Fu, S. G. Schirmer, and A. I.
Solomon, J. Phys. A {\bf 34}, 1679(2001).

\bibitem{wiseman94} H. W. Wiseman and G. J. Milburn,
Phys. Rev. Lett. {\bf 70}, 548(1993); H.M.Wiseman, Phys. Rev. A {\bf
49}, 2133(1994).


\bibitem{mancini98} S. Mancini, D. Vitali, and P. Tombesi, Phys. Rev. Lett. {\bf 80},
688(1998).

\bibitem{belavkin99} V. P. Belavkin, Rep. Math. Phys. {\bf 43}, 405(1999).

\bibitem{doherty99} A. C. Doherty and K. Jacobs, Phys. Rev. A {\bf 60}, 2700(1999).

\bibitem{carvalho07} A. R. R. Carvalho, J. J. Hope, Phys. Rev. A {\bf 76}, 010301(R)(2007).

\bibitem{yi08} L. C. Wang, X. L. Huang, and X. X. Yi, Phys. Rev. A {\bf 78}, 052112
(2008); H. Y. Sun, P. L. Shu, C. Li, and X. X. Yi,  Phys. Rev. A
{\bf 79}, 022119 (2009).

\bibitem{wiseman95} H. M. Wiseman,
Phys.\ Rev.\ Lett. {\bf 75}, 4587(1995).

\bibitem{berry01} D. W. Berry, H. M. Wiseman,and J. K. Breslin,
Phys. Rev. A {\bf 63} 053804,(2001).

\bibitem{jacobs07} K. Jacobs,
Quant. Information Comp.{\bf 7}, 127(2007).

\bibitem{geremia04a} J. M. Geremia,
Phys. Rev. A {\bf 70}, 062303(2004).

\bibitem{ahn02} C. Ahn, A. C. Doherty, and A. J. Landahl,
Phys. Rev. A {\bf 65}, 042301(2002).

\bibitem{hopkins03}  A. Hopkins, K. Jacobs, S. Habib, and K. Schwab,
Phys. Rev. B {\bf 68}, 235328(2003).

\bibitem{sarovar04} M. Sarovar,  C. Ahn, K. Jacobs, and G. J. Milburn,
Phys. Rev.  A {\bf 69}, 052324 (2004).

\bibitem{steck04} D. A. Steck, K. Jacobs,  H. Mabuchi,  T. Bhattacharya,
and S. Habib, Phys. Rev. Lett.{\bf 92}, 223004(2004).


\bibitem{geremia04b}M. A. Armen, J. K. Au, J. K. Stockton, A. C. Doherty,
and H. Mabuchi, Phys. Rev. Lett. {\bf 89}, 133602 (2002).


\bibitem{vettori02} P. Vettori, in Proceedings of the MTNS
Conferene, 2002.

\bibitem{ferrante02} A. Ferrante, M. Pavon, and G. Raccanelli, in
Proceedings of the 41st IEEE Conference on Decision and Control,
2002.

\bibitem{grivopoulos03} S. Grivopoulos and B. Bamieh, in Proceedings
of the 42nd IEEE Conference on Decision and Control, 2003.

\bibitem{mirrahimi04} M. Mirrahimi and P. Rouchon, in Proceedings of
IFAC Symposium LOLCOS 2004; In Proceedings of the International
Symposium MTNS 2004.

\bibitem{mirrahimi05} M. Mirrahimi and G. Turinici, Automatica {\bf
41}, 1987(2005).

\bibitem{altafini07} C. Altafini, Quantum Information Processing {\bf 6}, 9(2007).

\bibitem{wang08} X. Wang and S. Schrimer, arXiv: 0801.0702;
arXiv: 0901.4515.

\bibitem{yi09}X. X. Yi, X. L. Huang, Chunfeng Wu, and C. H. Oh, arXiv:0908.1048,
Phys. Rev. A, accepted.

\end{references}
\end{document}